\documentclass[twocolumn]{aastex631}

\usepackage{graphicx}	
\usepackage{amsmath}	
\usepackage{gensymb}

\newcommand{\bb}{\textbf{b}}
\newcommand{\bvv}{\textbf{v}}

\begin{document}


\title{Emergence of two inertial sub-ranges in solar wind turbulence: \\ dependence on heliospheric distance and solar activity}

\author[0009-0001-4841-1103]{Shiladittya Mondal}
\affiliation{Department of Physics, Indian Institute of Technology Kanpur, Kanpur 208016, India}

\author[0000-0002-3746-0989]{Supratik Banerjee}
\affiliation{Department of Physics, Indian Institute of Technology Kanpur, Kanpur 208016, India}

\author[0000-0002-5981-7758]{Luca Sorriso-Valvo}
\affiliation{Institute for Plasma Science and Technology (ISTP), CNR, Bari, Italy}
\affiliation{Space and Plasma Physics, School of Electrical Engineering and Computer Science, KTH Royal Institute of Technology, Stockholm, Sweden}



\begin{abstract}

The solar wind is highly turbulent, and intermittency effects are observed for fluctuations within the inertial range. By analyzing magnetic field spectra and fourth-order moments, we perform a comparative study of turbulence and intermittency in different types of solar wind measured during periods of solar minima and a maximum. 
Using eight fast solar wind intervals measured during solar minima between 0.3 au and 3.16 au, we found a clear signature of two inertial sub-ranges with $f^{-3/2}$ and $f^{-5/3}$ power laws in the magnetic power spectra. 
The intermittency, measured through the scaling law of the kurtosis of magnetic field fluctuations, further confirms the existence of two different power laws separated by a clear break. 
A systematic study on the evolution of the said sub-ranges as a function of heliospheric distance shows correlation of the break scale with both the turbulence outer scale and the typical ion scales. 
During solar maximum, on the contrary, the two sub-ranges are not omnipresent, thus showing more variability in the power spectra and intermittency scaling properties.

\end{abstract}



\section{Introduction}

The solar wind is the most accessible natural laboratory for studying space plasma turbulence \citep{BrunoCarbone2013}. Well above the ion inertial scale, the turbulent fluctuations of velocity $(\bvv)$ and magnetic field $(\textbf{B})$ can be described in the framework of magnetohydrodynamics \citep[MHD,][]{biskamp_2003}.
The highly irregular and dynamical structure of the solar corona, along with the supersonic and super-Alfvénic solar wind speed, entails strong nonlinear couplings that evolve into a turbulent cascade of energy corresponding to a power-law behaviour of the energy power spectral density $PSD(k)$ \citep[where $k$ is a wave vector][]{Coleman1968,frisch1995}. 
A $PSD(k) \sim k^{-5/3}$, observed universally in solar wind turbulence \citep{tu1995mhd, BrunoCarbone2013, Deepali21, Telloni2024}, is consistent with the prediction of the Kolmogorov phenomenology of a self-similar energy cascade within the so-called inertial range \citep{Kolmogorov1941}. 
In physical space, such universal cascade is manifested in terms of the linear scaling laws for the third-order moments of velocity and magnetic field fluctuations, which are also broadly observed in space plasmas \citep{Politano1998,lsorriso2007,banerjee2016,marino2023}. 
As typical in most turbulent flows, the energy cascade is not exactly self-similar, meaning that the scale-invariance of the energy transfer is only valid globally. Instead, strong local inhomogeneities arise in the flow, which is known as inertial-range intermittency \citep{frisch1995}. 
Indeed, nonlinear interactions generate small-scale coherent structures, such as vortices, current sheets, etc., that do not fill the available space in a self-similar way nor are randomly distributed, but rather tend to form inhomogeneously distributed clusters of bursts \citep{kolmogorov_1962,anselmet_gagne_hopfinger_antonia_1984,frisch1995}.
As a result, the statistical properties of the scale-dependent field fluctuations change with the scale. 
In particular, the probability distribution functions (PDFs) of the scale-dependent fluctuations progressively changes from a roughly Gaussian shape at large scale to a high-tailed distribution as the scale decreases \citep{tu1995mhd, lsorriso1999}.
This can be conveniently measured using the scaling laws of the high-order moments of the fluctuations' PDFs \citep{frisch1995,  carbone1996, biskamp_2003, banerjee2014}.
A standard example is the kurtosis $K$ (the normalised fourth-order moment, see Section \ref{sec:methods}), which for intermittent turbulence has a power-law scaling in the inertial range \citep{SorrisoValvo2015}.

The highly dynamic solar activity and the diversity of the originating regions produce solar wind with a variety of characteristics, the most evident being the plasma speed. 
While the fast solar wind (FSW, $>550 \ km \ s^{-1}$) mainly emanates from the polar coronal holes, the slow solar wind (SSW, $<400 \ km \ s^{-1}$) is believed to be originated from equatorial streamers \citep{belcherdavis1971, smith1978, phillips1995}. 
During high solar activity, however, both FSW and SSW are distributed at all latitudes instead of being confined exclusively to polar and equatorial regions, respectively. 
An interesting feature of FSW is the high Alfv\'enicity, namely the high correlation (or anti-correlation) between fluctuations in velocity and magnetic field $\bb=\mathbf{B}/\sqrt{\mu_0\rho}$ (where the normalization using the mass density $\rho$ is used to convert magnetic field into velocity units). In contrast, the SSW predominantly comprises of weak ${\bf v}$-${\bf b}$ correlations. 
This one-to-one correspondence, however, does not strictly hold during high solar activity, as a third type of wind is also observed. 
This wind, termed as Alfv{\'e}nic slow solar wind (ASSW), has low speed but is surprisingly permeated with high Alfv{\'e}nicity \citep{Marsch1981, amicis2011, damicis2015}. 
The degree of Alfv{\'e}nicity correlates with the nature of turbulence in different types of solar winds \citep[see][and references therein]{BrunoCarbone2013}. 
A high degree of Alfv{\'e}nicity corresponds to an imbalance between the inward and outward Alfvénic fluctuations of solar origin that propagate along the heliospheric magnetic field, which, according to MHD models, results in reduced nonlinear interactions and thus less developed turbulence \citep{kraichnan1965, BrunoCarbone2013}. 
On the contrary, low Alfv{\'e}nicity corresponds to comparatively more developed turbulence, owing to the stronger nonlinear interactions between balanced inward and outward perturbations. 
In solar wind streams, a broader $k^{-5/3}$ energy power spectrum is therefore observed for the slow wind, whereas a comparatively shorter spectrum is observed in the Alfv{\'e}nic fast and slow winds \citep{brunocarbone2005, BrunoCarbone2013, damicis2018,damicis2022}.


While the solar wind expands and accelerates through the heliosphere, the turbulence becomes more developed, with the fluctuations being majorly energized by the nonlinear interactions between the oppositely propagating Alfv{\'e}n fluctuations \citep{chandran2018}, switchbacks \citep{bale2021, sakshee2022}, large-scale structures, and instabilities \citep{bavassano1982b, roberts1992}. 
Using \textit{in-situ} spacecraft data, a steepening of the magnetic power spectra was found with increasing heliospheric distance, $R$, in the inner-heliosphere and beyond $1$ au \citep{bavassano1982a,roberts2010}. 
In addition, a decrease in \bvv-\bb~correlations and a broadening of the inertial range was also found as $R$ increases \citep{bavassano1998, bavassano1982a, davis2023}.
Recently, using high resolution data of the Parker Solar Probe it has been suggested that the magnetic spectral index evolves from a shallower $-3/2$ near the Sun (as close as 0.17 au), typical of strongly Alfvénic MHD turbulence to a more developed $-5/3$ at 1 au \citep{alberti2020, chen2020, shi2021, Sioulas2023b}. 
These observations are consistent with the idea of radial evolution of solar wind turbulence into more developed states and the corresponding non-adiabatic heating of the medium with increasing heliospheric distance \citep{marsch1982, cranmer2009, hellinger2011}. 
In addition, a power law behaviour for the kurtosis \citep{bruno2003,dimare2019, Carbone21,Hernandez2021} and an increase in intermittency in solar wind turbulence have been observed at greater heliospheric distances \citep{Sioulas2022,lsorriso2023}. 

Recently, more careful analysis of scaling laws are emerging that challenge this well established framework.
Preliminary studies suggest that a break in both the spectral density and the higher-order structure functions seems to characterize strongly Alfvénic solar wind intervals \citep{Wicks2011,Wu2022,telloni2022,sioulas2023a,Wu2023,lsorriso2023}. 
This would introduce a substantial modification of the scaling invariance of the solar wind MHD fluctuations.
However, the nature of such break and its implications on the dynamics of the solar wind turbulence have not been investigated in detail yet.


In this paper, we revisit the aforementioned problem and carry out a systematic study to provide a detailed description of the break observed in the solar wind scaling laws of turbulent fluctuations. 
Using \textit{in-situ} data of Helios and Ulysses during solar minima, we show that the break observed in spectra and kurtosis consistently separate two inertial sub-regimes, having $-3/2$ and $-5/3$ spectral indices, and weaker and stronger intermittency, respectively. 
In addition, we also study the radial evolution of the break scale to characterise the solar wind turbulence as a function of the heliospheric distance. 
During a solar maximum, a comparative study of FSW, ASSW and SSW measured by Ulysses shows a less clear classification of the inertial range, with breaks still being mostly present in Alfvénic streams.
In Sections \ref{sec:data} and \ref{sec:methods}, we briefly describe the data and methodologies used for the analysis. Section \ref{sec:results} provides the results obtained in our study during solar minima (\ref{sec:solarminm}) and maxima (\ref{sec: results solarmax}), respectively. Finally, in Section \ref{sec:conclusion}, we summarize our findings and conclude.

\section{DATA SELECTION}
\label{sec:data}

\begin{table}
 \caption{Intervals of FSW, ASSW and SSW used in our study. The intervals A3, A7, A8 (Helios-2) and A6 (Helios-1) are mentioned in \cite{dperrone2018} as well. The other intervals with abbreviations F\# (fast), S\small{\#} (slow), AS\# (Alfv\'{e}nic-slow) are from Ulysses spacecraft.}
 \label{tab:windsamples}
 \begin{tabular*}{\columnwidth}{c@{\hspace*{6pt}}|c@{\hspace*{8pt}}c@{\hspace*{8pt}}c@{\hspace*{8pt}}c@{\hspace*{8pt}}c}
 \hline
  Label & Year & Time Interval & $V_{sw}$ & R & lat\\[2pt]
        &      &    (MM-DD-HH)   &    (km/sec)   &         (au)    & ($\degree$)     \\[2pt]
  \hline 
  \multicolumn{5}{c}{Solar minimum} \\
  \hline \\
  A8 & 1976 & 04-14-14 -- 04-22-01 & 728.9 & 0.30 & -\\[2pt]
  A6 & 1976 & 03-14-10 -- 03-19-13 & 624.3 & 0.41 & -\\[2pt]
  A7 & 1976 & 03-15-18 -- 03-18-03 & 620.9 & 0.65 & -\\[2pt]
  A3 & 1976 & 01-21-21 -- 01-25-10 & 633.1 & 0.98 & -\\[2pt]
  F1 & 1995 & 01-21-00 -- 01-27-16 & 745.7 & 1.44 & -36.4\\[2pt]
  F2 & 1995 & 08-09-19 -- 08-15-12 & 795.0 & 2.10 & -78.7\\[2pt]
  F3 & 1995 & 11-12-00 -- 11-18-00 & 795.6 & 2.75 & -57.2\\[2pt]
  F4 & 1996 & 01-16-00 -- 01-22-00 & 765.7 & 3.16 & 50.7\\[4pt]
  
  \hline 
  \multicolumn{5}{c}{Solar maximum} \\
  \hline \\
  F5 & 2001 & 08-16-02 -- 08-18-02 & 734.4 & 1.64 & 63.9\\[2pt]
  F6 & 2001 & 09-09-14 -- 09-11-14 & 753.9 & 1.80 & 74.3\\[2pt]
  F7 & 2001 & 08-26-10 -- 08-28-10 & 690.9 & 1.71 & 68.7\\[2pt]
  F8 & 2001 & 02-18-07 -- 02-20-07 & 626.6 & 1.69 & -56.2\\[2pt]
  F9 & 2001 & 03-13-00 -- 03-15-00 & 694.9 & 1.56 & -44.2\\[2pt]
  AS1 & 2001 & 02-08-20 -- 02-10-20 & 353.5 & 1.53 & 53.4\\[2pt]
  AS2 & 2001 & 05-01-15 -- 05-03-15& 385.9 & 1.35 & -6.9\\[2pt]
  AS3 & 2001 & 06-21-23 -- 06-23-23 & 387.3 & 1.41 & 34.8\\[2pt]
  AS4 & 2001 & 07-05-08 -- 07-07-08 & 413.7 & 1.34 & 0.9\\[2pt]
  AS5 & 2001 & 06-05-12 -- 06-07-12 & 400.0 & 1.47 & -34.1\\[2pt]
  S1 & 2001 & 07-27-14 -- 07-29-14 & 367.4 & 1.76 & -60.6\\[2pt]
  S2 & 2001 & 05-06-05 -- 05-08-05 & 347.9 & 1.36 & -10.5\\[2pt]
  S3 & 2001 & 06-29-04 -- 07-01-04 & 430.4 & 1.38 & 29.5\\[2pt]
  S4 & 2001 & 05-16-06 -- 05-18-06 & 300.7 & 1.43 & 39.1\\[2pt]
  S5 & 2001 & 03-29-12 -- 03-31-12 & 469.0 & 1.35 & 16.9\\[4pt]
  \hline 
  
 \end{tabular*}
\end{table}
\begin{figure*}
 \includegraphics[width=\linewidth]{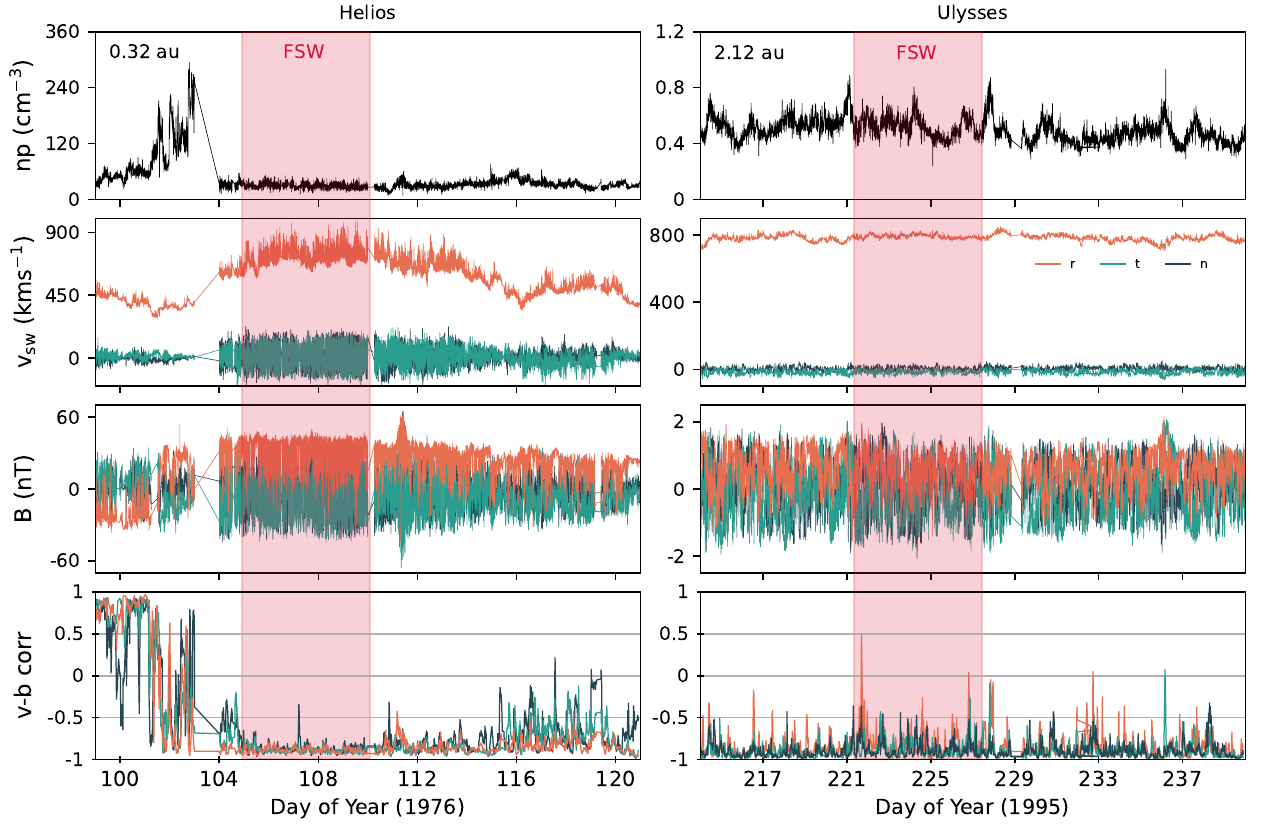}
 \caption{FSW intervals indicated by red boxes using Helios (left) and Ulysses (right) data, during solar minima. Top to bottom: proton number density, solar wind speed, interplanetary magnetic field, correlation coefficient between the components of proton velocity and magnetic field computed over a 2 hr window.}
 \label{fig1}
\end{figure*}
\begin{figure*}
 \includegraphics[width=\linewidth]{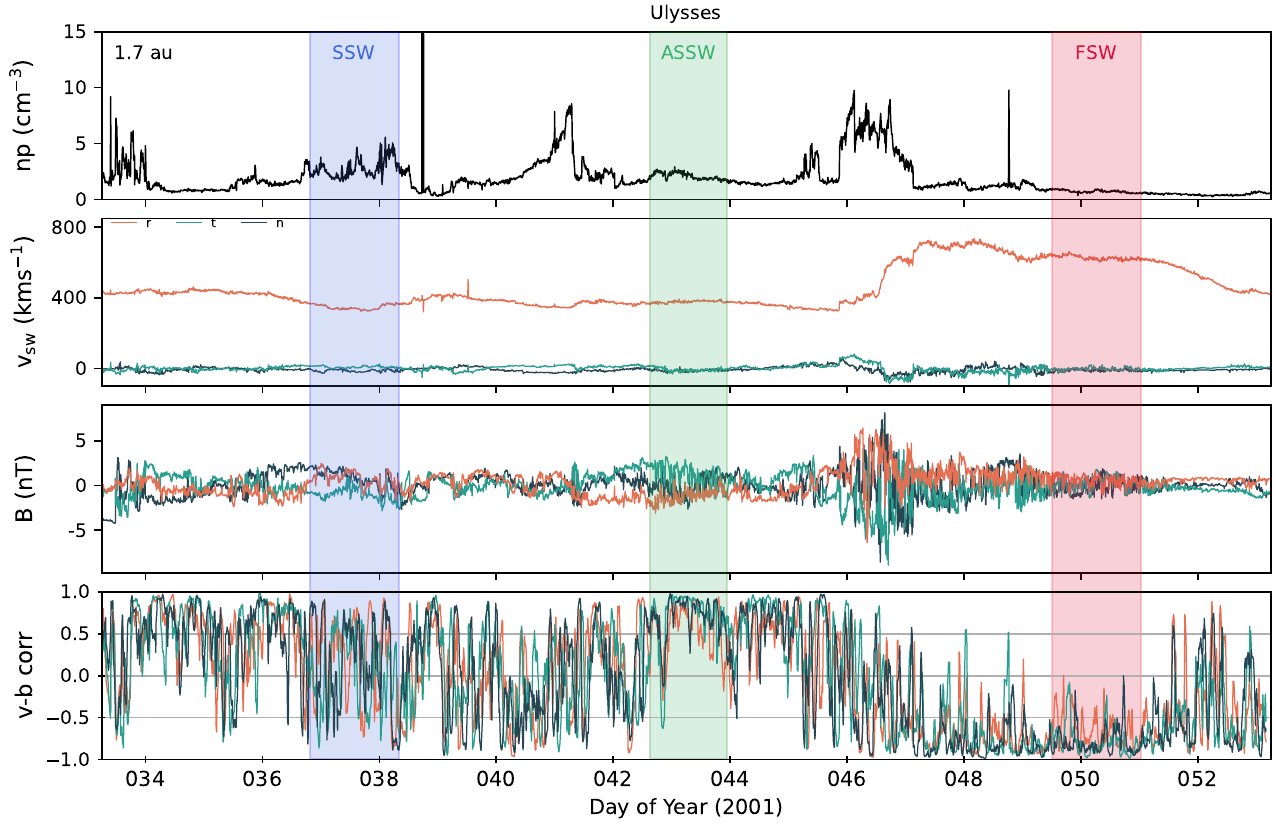}
 \caption{Different Ulysses intervals of SW during a period of solar maximum. Top to bottom: proton number density, solar wind speed, interplanetary magnetic field, correlation co-efficient between the components of proton velocity and magnetic field computed over a 2 hr window. Blue, green and red boxes represent SSW, ASSW, and FSW intervals, respectively.}
 \label{fig2}
\end{figure*}
%

For our analysis, we have used \textit{in-situ} data from the Helios and Ulysses spacecraft data repository publicly available at NASA CDAWeb (\href{https://cdaweb.gsfc.nasa.gov/}{https://cdaweb.gsfc.nasa.gov}) and AMDA science analysis system (\href{https://amda.irap.omp.eu/}{https://amda.irap.omp.eu}).
The plasma data for Helios and Ulysses have been obtained from the E1 Plasma Experiment instrument and the Solar Wind Observations Over the Poles of the Sun (SWOOPS) instrument, respectively. For magnetic power spectrum and kurtosis scaling, we use 6 s resolution magnetic-field data from the E3 Flux-gate Magnetometer (FGM) onboard Helios and 1 s resolution magnetic field data from the Vector Helium Magnetometer (VHM) onboard Ulysses spacecraft. 
During a declining phase of solar activity near a solar minimum between 1975 and 1976, Helios 1 and 2 recorded several streams of FSW from a coronal hole (or the same source), which sustained through nearly two solar rotations \citep{bruno2003}. 
Several intervals of the fast wind expelled from this coronal hole were also identified by \cite{dperrone2018}. 
In particular, for our current analysis, we use the streams A3, A6, A7, and A8, ranging from 0.3 au to 1 au, mentioned therein.
Each chosen interval contains negligibly small amount of data gaps, is free of any considerable mean trend, and turns up to be reasonably stationary. 
The stationarity is assured by the approximate constant average of sub-intervals of different lengths. Extending our analysis beyond 1 au, we use four intervals of FSW at varying heliospheric distances (F1---F4 as listed in Table \ref{tab:windsamples}), recorded by Ulysses during the years 1995-1996. Typical features of certain FSW intervals in the inner and outer heliosphere used in our analysis with high ${\bf v}$-${\bf b}$  correlations are shown in Fig. \ref{fig1}.

In order to interpret our findings, we also need to compute the co-spectra of cross-helicity $\sigma_c$ (see Section \ref{sec:methods}), for which we have used the 40.5 s resolution magnetic field and proton velocity data from the E3 FGM and the E1 Plasma Experiment instrument onboard Helios. We use degraded resolution for the magnetic field data in order to keep coherence with the available plasma data from the data repository. 
A similar analysis cannot be done using the plasma data of Ulysses where the data resolution is 240 s, and hence cannot be used to capture the required length scales of our interest. 

During solar maximum, five Ulysses intervals each for the three types of solar wind were selected following similar methods prescribed in \cite{damicis2018} based on their speed, proton density, and Alfv{\'e}nic correlations (see Table \ref{tab:windsamples}). 
A particular case study shown in Fig. \ref{fig2} represents several properties of the different types of wind within a 20-day interval. While ASSW looks very similar to SSW with respect to the flow speed ($< 400$ km/sec), it is characterised by low proton density ($\sim 1$ particle/cm$^\text{3}$) and high Alfv\'{e}nicity ($\sim 0.6$) similar to FSW. 
These findings are in agreement with previous studies \citep{belcherdavis1971, Marsch1981, damicis2015}.

\section{ANALYSIS METHOD}
\label{sec:methods}

Our analysis is mainly based on the computation of (i) the magnetic power spectral density ($PSD$), (ii) the kurtosis ($K$) or the normalized fourth-order moments of magnetic field fluctuations, and (iii) the cross-helicity co-spectra ($\hat{\sigma_c}$). 
All of the data sets were made evenly sampled by interpolating the data gaps before any of the computations. 

Since all the intervals used in our study are permeated by super-Alfv\'enic solar wind, one can practically use Taylor's hypothesis, which means if the phase speed of the fluctuations is much smaller than the flow speed of the solar wind, the fluctuations can be considered as frozen (or slowly evolving) as the flow sweeps the probe \citep{taylor1938}. 
When using single-point measurements in the form of a time series, the only accessible direction for the increments is along the bulk flow.
This provides an equivalence between the longitudinal (along the flow) length scale $\ell$ and the corresponding time scale $\tau$ as ${\ell} = V_{sw} \tau$, where $V_{sw}$ is the mean solar wind speed.
Therefore, we define the increments of the $i^{th}$ component (with $i=r,t,n$ indicating the vector components in the standard RTN coordinate system) of the magnetic field as $\Delta B_i(t,\tau)=B_i(t+\tau)-B_i(t)$.
In order to capture both magnitudinal and directional fluctuations of $\mathbf{B}$, we define the $n^{th}$ order structure function as:

\begin{equation}\label{eqn1}
    S_{n}(\tau) = \Biggl \langle \left[ \sum_i \left(\Delta B_i \right)^2 \right]^{n/2} \Biggr \rangle,
\end{equation}
where $\langle \cdot \rangle$ represents the ensemble average \citep{bruno2003}.
The corresponding kurtosis (K) is then calculated using the standard expression:
\begin{equation}\label{eqn2}
    K(\tau) = \frac{S_{4}(\tau)}{[S_{2}(\tau)]^2}. 
\end{equation}
Note that, when each $\Delta B_i$ follows a Gaussian distribution with zero mean, $K (\tau)$ is equal to $5/3$ (see appendix Section \ref{AA}). For a self-similar, non intermittent flow, in the inertial range of scales (namely much smaller than the energy-injection scales and larger than the dissipative scales) the $n^{th}$ order structure function is expected to scale as $S_n(\tau) \propto \tau^{np}$, where $p$ is a phenomenological constant \citep{frisch1995}. 
It is therefore straightforward to see that $K$ becomes independent of $\tau$. 
However, in the presence of intermittency, this linear scaling does not hold and the simplest intermittency model can be given as $S_n(\tau) \propto \tau^{np + q{(n)}}$, where $q{(n)}$ is a nonlinear correction accounting for the intermittent structures. 
For the kurtosis, this leads to a power-law scaling $K(\tau) \sim \tau^{-\kappa}$, with $\kappa=q{(4)}/2q{(2)}$. 
Such a scaling, universally observed in fluid turbulence, has recently been quantitatively described in the case of solar wind turbulence as well \citep{dimare2019, Hernandez2021, lsorriso2023}. 
In this work, we study the scaling properties of $K$ of the magnetic field fluctuations at different heliospheric distances.

Finally, the magnetic energy spectra and normalized cross-helicity co-spectra are defined as $PSD = \hat{B_i}^\dagger \hat{B_i}$ and $\hat{\sigma_c} = (\hat{b_i}^\dagger\hat{v_i} + \hat{v_i}^\dagger\hat{b_i})/(|\hat{b_i}|^2 + |\hat{v_i}|^2)$ respectively, where $\hat{B_i}$, $\hat{b_i}$ and $\hat{v_i}$ are the Fourier transforms of $B_i$, $b_i$ and $v_i$, respectively, with summation being intended over the repeated indices (where $i=r,t,n$).

\section{RESULTS AND DISCUSSIONS}
\label{sec:results}

\subsection{Observations during Solar Minimum}\label{sec:solarminm}
\begin{figure*}
 \includegraphics[width=\linewidth]{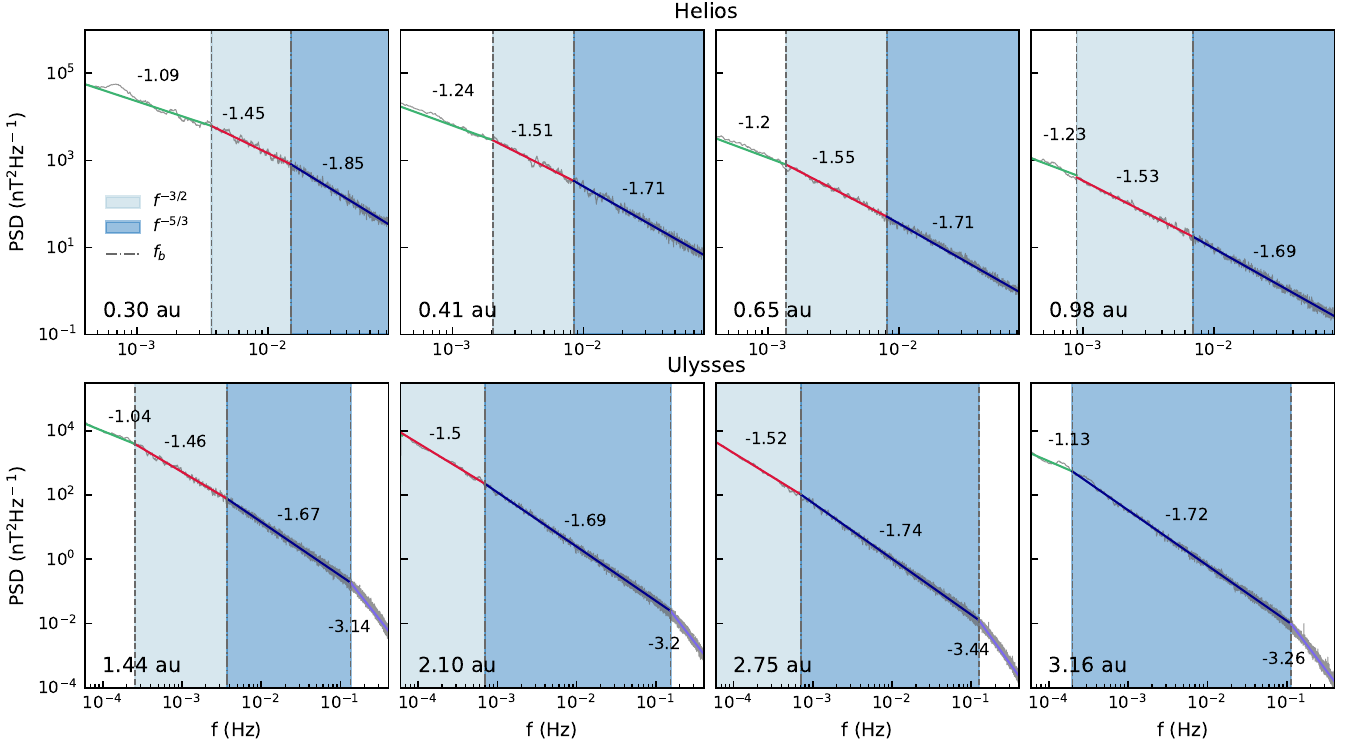}
 \caption{Magnetic power spectral trace of FSW intervals (smoothed using a running mean window) at varying heliospheric distances $R$ during solar minima. Top panels: Helios data (year 1976), bottom panels: Ulysses data (years 1995-1996). In all panels, the distance of the interval from the Sun is indicated. Vertical lines indicate the $f^{-1}$ break (dashed), the newly observed break $f_b$ (dot-dashed, separating the light and deep blue areas jointly forming the traditional inertial range), and the ion-scale break (Ulysses only). In each range, a power-law fit is shown (coloured lines) along with the corresponding spectral exponent.}
 \label{fig3}
\end{figure*}
\begin{figure*}
    \centering
    \includegraphics[width=1\linewidth]{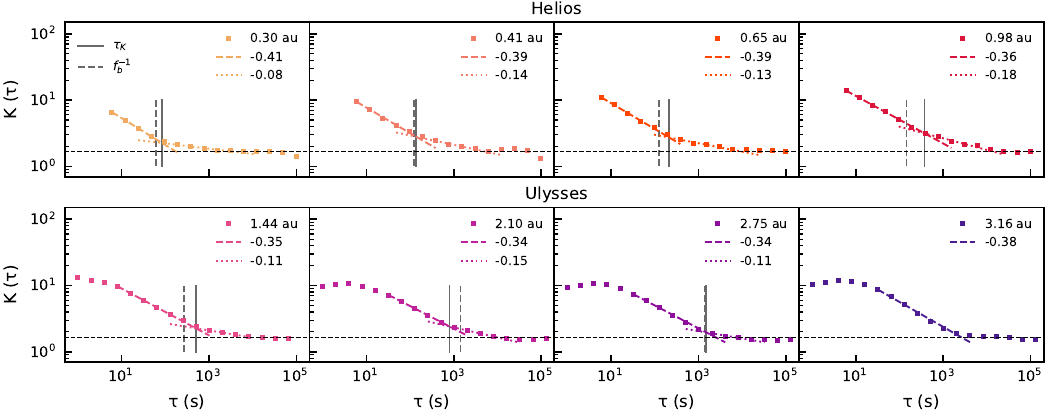}
    \caption{Kurtosis $K(\tau)$ of magnetic field fluctuations for several intervals of FSW during periods of solar minima. Top panels: Helios data in the inner heliosphere from a sustained coronal hole. Bottom panels: Ulysses data in the outer heliosphere at varying distances and latitudes (the distance of each interval is indicated and associated with a given colour).
    Power-law fits and the corresponding scaling exponents are indicated. Vertical lines indicate the observed break, $\tau_K$ (solid lines), and the timescale corresponding to the spectral break, $1/f_b$ (dashed).}
    \label{fig4}
\end{figure*}
\begin{figure}
 \includegraphics[width=1\linewidth]{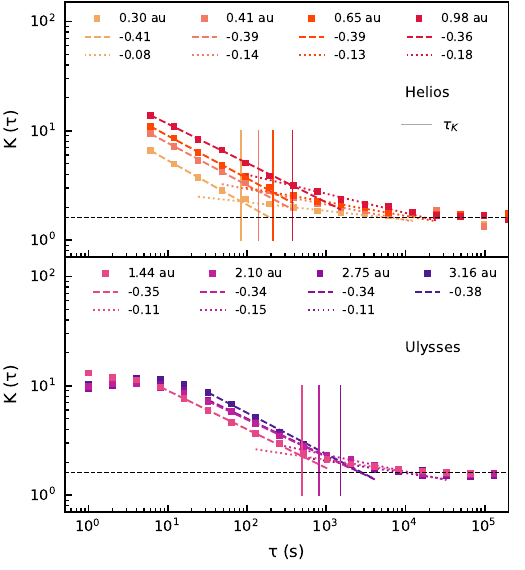}
 \caption{Consolidated plot of the kurtosis $K(\tau)$ scalings of the FSW intervals during solar minima. Top panel: Helios data ($R<1$ au); Bottom: Ulysses data ($R>1$ au). Vertical full lines indicate the break, $\tau_K$, shifting towards larger scales with $R$.}
 \label{fig5}
\end{figure}
During a period of solar minimum in 1976, using data from Helios spacecraft, we study FSW streams in the inner heliosphere (at $0.3, 0.41, 0.65,$ and $ 0.98$ au) from a sustained coronal hole near the ecliptic plane. 
Beyond $1$ au, FSW streams are studied using Ulysses data collected during the 1995-1996 solar minimum at varying heliospheric distances (at $1.44, 2.1, 2.75,$ and $ 3.16$ au), which were also measured at different latitudes.
%
%
In Fig. \ref{fig3}, we have drawn the magnetic power spectral traces, smoothed using a running mean window.
Top panels refer to Helios intervals, while bottom panels to Ulysses. 
As typically observed in the Alfv\'enic solar wind, at low frequencies we can identify a large-scale, energy-containing range (white background in the figure), where the power decays as $\sim f^{-1}$. Fitted power laws and the corresponding scaling exponents are shown as green lines. 
A break identifies a clear change in the power-law scaling exponent, as indicated by vertical dashed lines. 
Such break can be associated with the correlation scale of the turbulence. 
The low-frequency range is clearly visible in Helios data, while it is only indicatively present in the Ulysses intervals. This is consistent with the well-known shift of the correlation scale towards lower frequency with increasing $R$ in the solar wind \citep{davis2023}.
The $f^{-1}$ range is followed by the usual inertial range of turbulence, where the spectrum roughly follows an $f^{-5/3}$ power law dependence \citep{brunocarbone2005, BrunoCarbone2013}.
However, a more accurate inspection shows that a further break emerges within such range, indicated by the vertical dot-dashed lines separating the light and deeper blue shaded areas in Fig. \ref{fig3}. 
Although the dynamical range of frequencies is relatively short, for intervals other than that at $3.16$ au, it is possible to identify two different sub-ranges with different power laws as demonstrated by the red and blue lines, with the associated scaling exponents indicated nearby. 
In the lower-frequency range (light blue background), the spectral index approaches ${-3/2}$, whereas at larger frequencies (deep blue background), the spectra show a transition to a ${-5/3}$ spectral index usually observed in non-Alfv\'enic solar wind \citep{brunocarbone2005,BrunoCarbone2013, alexandrova2009, damicis2018}.
In isotropic turbulence, whereas an $f^{-5/3}$ scaling often represents an energy cascade by eddy fragmentation in strong turbulence, $f^{-3/2}$ scaling can possibly be explained by an energy cascade through the sporadic interaction of Alfv\'{e}nic wave packets in MHD turbulence \citep{Kolmogorov1941, iroshnikov1963, kraichnan1965}.
However, $-5/3$ and $-3/2$ power laws can also be obtained under various circumstances if anisotropy is taken into account \citep{goldreichsridhar1995, goldreichsridhar1997, boldyrev2006, chandran2015}.
Irrespective of the true nature of energy cascade, a single power law is often assumed for the magnetic power spectra in the frequency range $10^{-4} - 10^{-1}$ Hz \citep{brunocarbone2005,BrunoCarbone2013}, 
although a few studies have found variation in the power law exponents in the inertial range of magnetic power spectra \citep{Wicks2011,sioulas2023a,damicis2025} as well as the scaling of higher order structure functions \citep{Wu2022,lsorriso2023,damicis2025}.
In our study, the co-existence of the two sub-regimes (with $-3/2$ and $-5/3$ spectral indices) within the turbulence spectra of FSW has been consistently observed at various heliospheric distances both in the inner as well as the outer heliosphere.
The break scale between those two sub-ranges, $f_b$, appears to shift towards lower frequencies (approaching the correlation scale) with increasing heliospheric distance. 
This is consistent with the fact that, when using single power law, a ${-3/2}$ scaling has been observed for solar wind close to the Sun, whereas a steeper ${-5/3}$ power law is obtained at and beyond $1$ au \citep{chen2020, shi2021,Sioulas2023b}.
Finally, in the Ulysses intervals, the ion-scale breaks are visible, separating the MHD range from the sub-ion range, where Hall effects and other kinetic effects start to affect the cascade (white background) \citep{banerjee2017,halder2023}. 
Such break is usually observed at frequencies $\sim 10^{-1}$ Hz, which is the upper cut-off for the MHD range.
However, similar breaks do not turn up in the Helios intervals due to the low cadence of the data used here.
%
%

To further investigate on the sub-inertial range spectral break, $f_b$, we study the kurtosis $K(\tau)$ for all of the eight FSW intervals.
The scaling of $K(\tau)$ defined in Section \ref{sec:methods} are depicted for Helios and Ulysses data in Fig. \ref{fig4} top and bottom panels, respectively, for each $R$.
To inspect on the general radial trend of intermittency, we have drawn a consolidated plot for the Helios and Ulysses intervals (see Fig. \ref{fig5}). From this figure one can conclude that the value of $K$ at all scales increases with increasing $R$, thus implying higher intermittency with increasing heliospheric distance, in agreement with previous studies \citep{bruno2003, lsorriso2023, Sioulas2023b}.
At each given distance $R$, $K$ is systematically found to decrease as one moves towards the larger scales. This is consistent with the notion that deviation from Gaussian statistics increases at smaller scales \citep{frisch1995,lsorriso1999}.
Upon reaching the typical correlation scales of the flow ($\tau \simeq 10^4$ s), corresponding to the $f^{-1}$ power law in energy spectrum (see Fig. \ref{fig3}), the kurtosis saturates to a constant value $K\simeq 1.67$, representing a quasi-Gaussian distribution (with a non-zero skewness) of the fluctuations of the magnetic field components (see Appendix \ref{AA}).
Within the inertial range, from the nature of $K(\tau)$ in Fig. \ref{fig4}, a clear signature of broken power law is observed. 
While two breaks are visible for Ulysses data (with $1$ s resolution), the small-scale break at around $\tau\sim 10$ s is missing for the intervals using Helios magnetic field data with $6$ s resolution. 
The other break which occurs at a larger $\tau$ (solid vertical lines) is clearly visible for both Helios and Ulysses data. 
As observed for the spectra, this break scale, $\tau_K$, shifts towards larger scales as $R$ increases.
Within the distance range of $0.3$---$2.75$ au, $\tau_K$ is found to increase from $\sim 100$ s to $\sim 1500$ s. 
It is to be emphasized here that except for certain cases, the appearance of the break $\tau_K$ is persistent in the component-wise $K$ scaling as well (see Figs. \ref{fig:app1} and \ref{fig:app2} in Appendix \ref{AB}).
A detailed list of the break scale $\tau_K$ as a function of $R$ is given in Table \ref{table2}. 
As it is evident from Fig.s \ref{fig4} and \ref{fig5}, $\tau_K$ separates the steeper power law ($K\sim\tau^{-\kappa}$ with $\kappa \simeq 0.37$ averaged over the eight intervals) at smaller scales (dashed lines) from the less steeper one ($\kappa \simeq 0.11$ on average) at large scales (dotted lines), but with an exception.
Note that for the Ulysses interval at $R=3.16$ au, $K(\tau)$ reaches the Gaussian regime without going through the large-scale break, suggesting that the turbulence has fully developed, which transforms the shallower scaling range at large scale into the steeper power law at smaller scales. 


\begin{table}
    \centering
    \begin{footnotesize}
    \caption{Variation of the break scales observed in Kurtosis ($K$) scaling, $\tau_K$, and in magnetic power spectra, $f_b$, as a function of heliospheric distance $R$.}
    \label{table2}
    \begin{tabular}{c|ccccccc}
        \hline
        $R$ (au)   & $0.3$ & $0.41$ & $0.65$ & $0.98$ & $1.44$ & $2.1$ & $2.75$\\[2pt]
        \hline
        $f_b$ (Hz) \tiny{$({\times10^{-3}})$} & $14$ & $8.2$ & $8$ & $6.9$ & $3.8$ & $0.7$ & $0.7$ \\[2pt]
        \hline
        $\tau_K$ (s) & $84$ & $140$ & $214$ & $374$ & $499$ & $799$ & $1510$ \\[2pt]
        \hline
    \end{tabular}
    \end{footnotesize}
\end{table}

As mentioned in the introduction, similar broken power-law behaviour for $K(\tau)$ in FSW has already been observed by \cite{lsorriso2023}. 
However, those authors suggested that $\tau_{K}$ might correspond to the break between low-frequency $f^{-1}$ regime to Kolmogorov $f^{-5/3}$ regime in the magnetic power spectra. 
This was inspired by the fact that $f^{-1}$ regime is exclusively found in FSW intervals, and the $f^{-1}$ break also shows nearly similar behaviour to $1/\tau_K$ as $R$ changes \citep{davis2023}. 
Instead, for all the intervals where the break is observed, it is systematically found in our study that $1/\tau_K$ occurs at a higher frequency (roughly by a factor $10$) than the $f^{-1}$ break scale (see Fig. \ref{fig3}). 
The inverse of $\tau_K$ is typically corresponding to $f_b$, although with some consistent small discrepancy that could be due to the different frequency response of Fourier transform and scale-dependent increments (see Fig. \ref{fig4} where both $\tau_K$, solid lines, and $1/f_b$, dashed lines, are drawn). 
The two scaling ranges in the kurtosis therefore approximately correspond to the two inertial sub-ranges observed in the spectrum. 
Since $PSD$ and kurtosis are related quantities, the observation of a double power law in both supports the robustness of the break and, therefore indicates the emergence of a new characteristic scale in the inertial range that marks the transition from $f^{-3/2}$ to $f^{-5/3}$ regime.

\begin{figure}
 \includegraphics[width=0.98\linewidth]{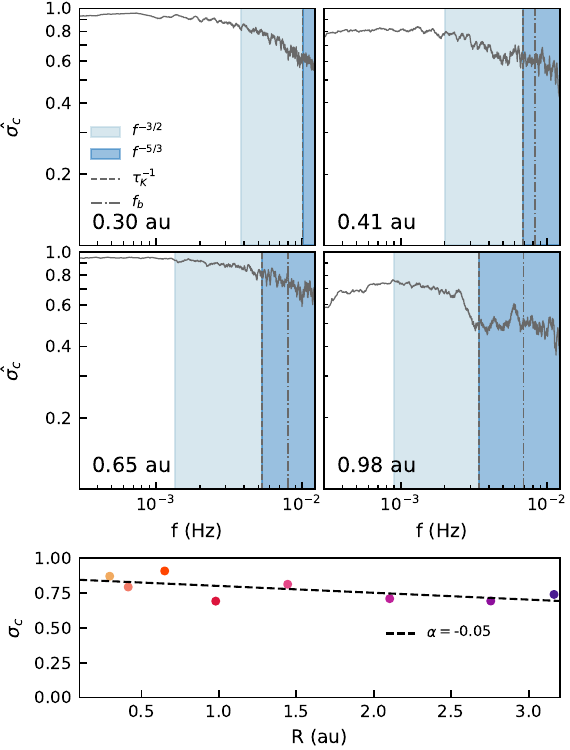}
 \caption{Top: Normalized cross-helicity spectrum $\hat{\sigma_c}$ (smoothed using a running mean window) of the four Helios FSW intervals in the inner heliosphere. The light and deep blue shaded regions depict the $-3/2$ and $-5/3$ regimes respectively. The spectral break frequency $f_b$ and the frequency associated with the kurtosis break, $\tau_K^{-1}$, are indicated by dot-dashed and dashed lines respectively. Bottom: Cross-helicity $\sigma_c$ of all the FSW intervals as a function of the heliospheric distance $R$.}
\label{fig6}
\end{figure}
\begin{figure}
 \includegraphics[width=\linewidth]{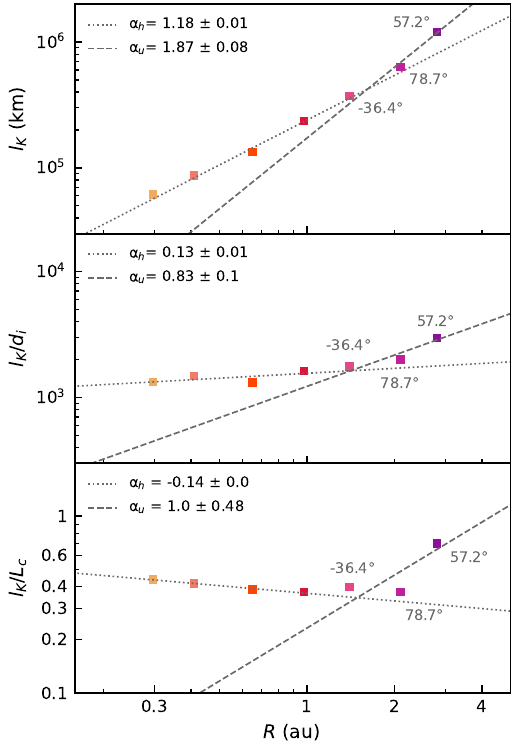}
 \caption{FSW during solar minima. Top: Kurtosis scaling break $l_{K}(=V_{sw}\tau_K )$ versus the heliospheric distance $R$. The same break scale $l_K$ normalized by the ion-inertial length scale $d_i$, $l_{K}/d_i$ (middle) and $l_K$ normalized by the correlation length $L_c$ (bottom) as a function of $R$.
 The different colours reproduce the colours in Fig. \ref{fig3}, and in the Ulysses intervals, the latitude is indicated. Two power laws were identified in the inner and outer heliosphere, respectively. The fitted power laws and the corresponding parameters are indicated. }
 \label{fig7}
\end{figure}

\begin{figure*}
 \includegraphics[width=\linewidth]{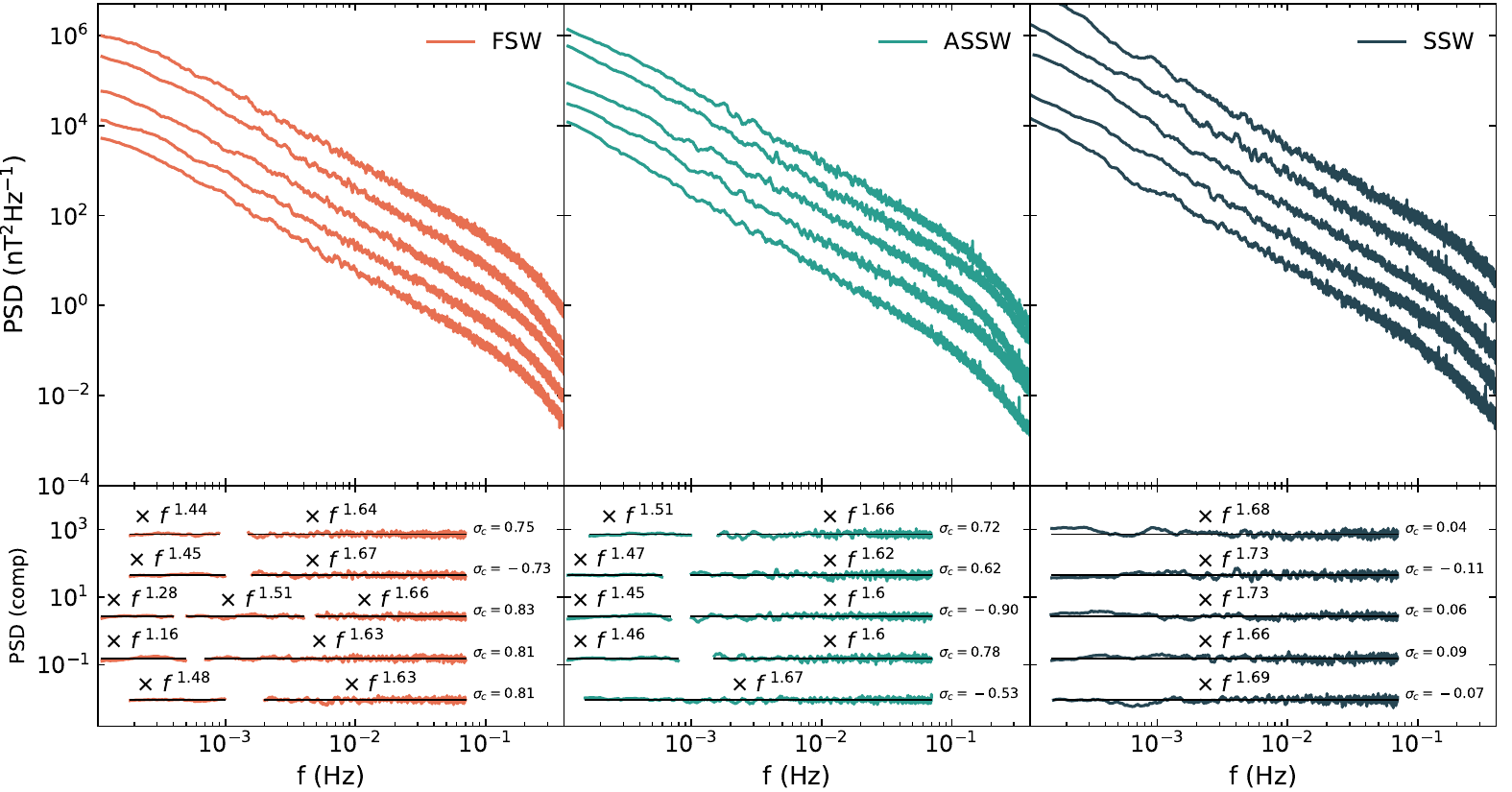}
 \caption{Magnetic field power spectral trace (smoothed using a running mean window) of the FSW, ASSW \& SSW intervals tabulated in Table-\ref{tab:windsamples} during solar maximum  (the PSD's are artificially shifted for representation). The SSW exhibits a broad inertial range while different power law breaks are present in the case of ASSW \& FSW. In each range, the compensated spectra is shown along with the corresponding spectral exponent and the normalized cross-helicity.}
 \label{fig9}
\end{figure*}
%
%
Summarizing, from the existence of the two turbulent inertial sub-regimes it is clear that as we move from the larger towards the smaller scales, the nature of turbulence also varies.
This variation becomes more apparent when we examine the cross-helicity spectrum for the FSW intervals within the inner heliosphere (Fig. \ref{fig6}, top panel). 
The same could not be computed for the FSW beyond $1$ au due to the limitation in terms of low plasma data resolution, as mentioned in Section \ref{sec:data}. 
Nevertheless, for all the FSW intervals in the inner heliosphere we see that the $\hat{\sigma_c}$ power decreases as we move from larger to smaller scales (see Fig. \ref{fig6}). 
Thus, with the forward progression of the turbulent cascade, the imbalance between the inward and outward Alfv\'en modes propagating along the mean magnetic field decreases to a more balanced state. 
Although there is no evidence of a well defined transition scale, these plots confirm the well known fact that, within the inertial range, larger scales are typically more unbalanced than smaller scales \citep{BrunoCarbone2013}. 
While recent studies have shown the transition from a weak to a strong turbulence regime on moving towards smaller scales \citep{zhao2024}, a transition from imbalanced ($|z^{+2}|\gg |z^{-2}|$, or vice-versa) to a balanced ($|z^{+2}|\sim|z^{-2}|$) turbulent state could as well be associated with the steepening of the spectra from the low frequency $f^{-3/2}$ regime to the higher frequency $f^{-5/3}$ regime.
A similar gradual change from an imbalanced towards a relatively balanced state is also evident with increasing heliospheric distance $R$. 
In the bottom panel of Fig. \ref{fig6}, we show the radial dependence of the normalized cross-helicity $\sigma_c=\langle \delta \mathbf{v}\cdot \delta \mathbf{b}\rangle / \langle|\delta \mathbf{v}|^2 + |\delta \mathbf{b}|^2\rangle$, where $\delta$ indicates fluctuations with respect to the interval mean and the average $\langle\cdot\rangle$ is done over the interval. Even though $\sigma_c$ shows large values associated with FSW, it declines slowly with the distance to the Sun $R$, as understood from the indicated linear fit with slope $\alpha=-0.05$.
This is again consistent with the absence of the $f^{-3/2}$ regime at $R=3.16$ au and recent observations of change in the inertial-range spectral index from $-3/2$ to $-5/3$ with increasing $R$ \citep{chen2020,shi2021, Sioulas2023b}.

We further determine the evolutionary nature of the break scale with radial distance, and investigate its relationship with the typical ion and correlation scales. 
In Fig. \ref{fig7} (top), we show the radial evolution of $\tau_K$, appearing in the scaling of $K$, converted from time scale to length scale, $l_{K}$, via Taylor's hypothesis as mentioned in Section \ref{sec:methods}. 
Clearly, $l_K$ shift towards larger scales with $R$, as evident from Fig. \ref{fig5} and Table \ref{table2} as well.
We see that a robust power-law relation exists between $R$ and $l_{K}$, with $l_{K}$ evolving as $l_{K} \propto R^{\text{ } 1.18}$ for $R<1$ au and $l_{K} \propto R^{\text{ } 1.87}$ for $R>1$ au.
The central panel in Fig. \ref{fig7} shows how the break scale behaves with $R$ when normalized to the ion-inertial length scale, $d_{i} = c/\omega_{pi}$ (where $\omega_{pi}$$ =\sqrt{ne^2/\epsilon_0 m}$ is the plasma frequency). 
The ion-inertial scale has been found to vary between $\sim45$ to $\sim500$ km for $R$ ranging from $R \simeq$ 0.3---3.2 au. 
After normalization, we find that the evolutionary nature is nearly lost for FSW intervals in the inner heliosphere near the ecliptic plane, with a residual weak $R^{\text{ } 0.13}$ dependence, and $l_{K}$ is $\sim10^3$ times $d_{i}$. 
A similar pattern was observed (but not shown) after normalization with the ion gyro-radius $\rho_{i} = v_{th}^\perp / \Omega_{i}$, in the inner heliosphere (the $\rho_{i}$ in the outer heliosphere could not be computed again due to data limitations).  
Note that the typical ion scales have an approximately linear radial increase up to 5 au \citep[e.g., see][]{BrunoTrenchi}, which might explain the constant radial trend of the normalized break scale.
However, beyond $1$ au, it is to be noted that even after normalization, the evolutionary nature of $l_{K}$ still persists so that only the radial trend of the break decouples from that of the ion scales. 
The residual power law could be associated with the variation in heliospheric latitude (and to the associated variation of the angle between the large-scale magnetic field $B$ and the bulk speed $V_{sw}$) at which the FSW streams were sampled, indicated in the labels in Fig. \ref{fig7}. 
Understanding this variation of $l_{K}$ with latitude and the angle between the bulk wind speed and the large-scale magnetic field would be interesting for a future study but is currently beyond the scope of this paper. 
In order to compare the break scale $l_K$ and the correlation scale $L_c$, we have drawn $l_K$ normalized to $L_c$ as a function of $R$ (see Fig. \ref{fig7} bottom). 
Here $L_c$ is the Taylor-shifted $\tau_c$, which is the time lag at which the trace of the correlation matrix of $\mathbf{B}$ decreases to $1/e$ of its initial value.
It is evident from the plot that, for $R<1$, a small power-law exponent is observed, $l_K/L_c \sim R^{\text{ } -0.14}$, so that the normalization to the correlation scale removes the radial dependence, similar to what we observe when normalized to the ion scale.
Moreover, in this case, $l_K$ is $\sim 0.4$ times $L_c$ and certainly does not correspond to scales within the $f^{-1}$ power law regime in the spectrum, contrary to what has been suggested previously \citep{lsorriso2023}.
For $R>1$, $l_K$ approaches $L_c$, thereby explaining the absence of the $f^{-3/2}$ regime in the $R=3.16$ au interval and supporting recent observations of spectral steepening of the inertial range with increasing $R$ \citep{chen2020, shi2021, Sioulas2023b}.
Note that, in the inner heliosphere, break scales normalized to both the characteristic ion scale and the correlation scale follow a comparably weak radial dependence of $R^{\text{ } 0.13}$ and $R^{\text{ } -0.14}$, respectively.

\begin{figure*}
 \includegraphics[width=\linewidth]{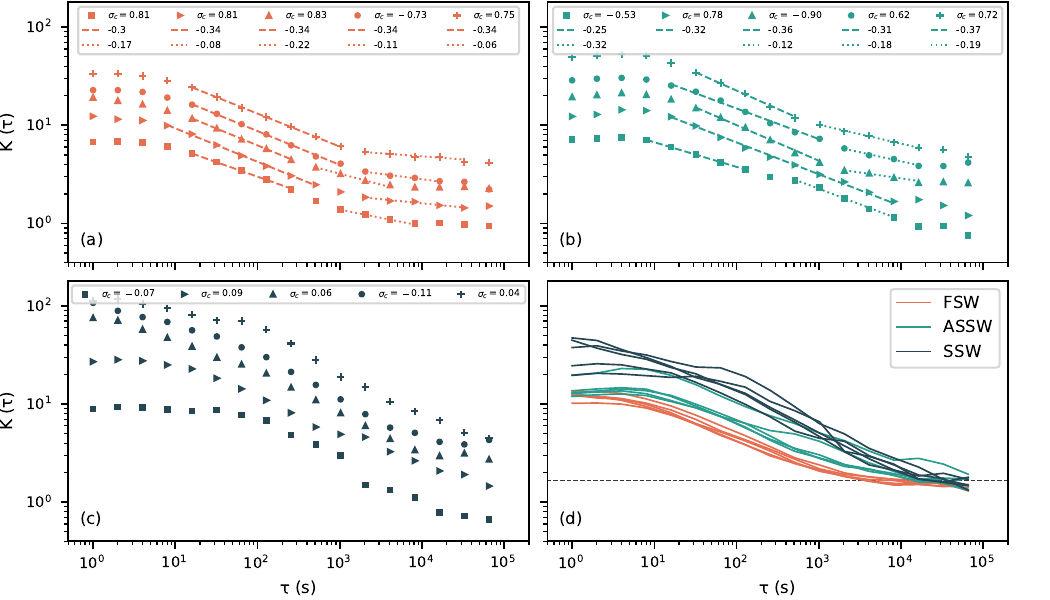}
 \caption{Kurtosis ($K$) of the magnetic field fluctuations as a function of time scale ($\tau$) for several intervals of (a) FSW (orange), (b) ASSW (green) and (c) SSW (deep blue), during solar maximum. The normalized cross-helicity is also indicated in the legend. 
 In panels (a), (b) and (c), the kurtosis for each interval have been artificially shifted for better representation while panel (d) shows a consolidated plot of all the intervals (with no artificial shifting) of the three types of solar wind.}
 \label{fig8}
\end{figure*}

\subsection{Observations during Solar Maximum}\label{sec: results solarmax}

We now perform a similar spectral and intermittency analysis using the set of intervals recorded during the solar maximum (see Table \ref{tab:windsamples}).
While the previous section was confined to only analyzing FSW, in this section, we take into consideration the three main solar wind types, namely FSW, SSW and ASSW.
Previous studies on spectra and intermittency mostly focused on FSW and SSW \citep{bruno2003, dimare2019, Carbone21, lsorriso2021}. 
More recently, the spectral properties of ASSW, which permeates the heliosphere during periods of high solar activity, were also examined \citep{damicis2021,damicis2022}. However, such studies did not include intermittency. Moreover, a comparative analysis between FSW, SSW and ASSW at solar maxima has not yet been conducted. 
Thus, in this section, we examine the intermittency properties of ASSW \citep{Marsch1981, amicis2011, damicis2015} in comparison with the other two types of wind using Ulysses data, during the ascending phase of solar cycle 23 (year 2001), at $R\simeq1.5$ au. 


In Fig. \ref{fig9}, the smoothed magnetic power spectra are shown for all intervals listed in Table \ref{tab:windsamples}. Compensated spectra in the relevant regions are shown in the bottom panels, where the values of the normalized cross-helicity are also indicated for each stream.
In contrast to the solar minima, where two regimes with $f^{-3/2}$ and $f^{-5/3}$ were systematically found, during solar maximum we find FSW intervals both with and without the $f^{-3/2}$ regime.
This could be due to the fact that the break between the $f^{-3/2}$ and $f^{-5/3}$ regime has evolved to larger scales beyond the correlation scale of turbulence, which was measured to be $\sim 1250$ s $(8\times10^{-4}$ Hz$)$ for these intervals. 
A similar observation was made in the previous section for the interval at $R=3.16$ au. 
Note also that a possible role of latitudinal variation (which spans from $-56^\circ$ to $75^\circ$ with respect to the ecliptic, see Table \ref{tab:windsamples}), was not apparent but cannot be excluded.
For the ASSW, the $f^{-3/2}$ regime is also evident in the spectra for most of the intervals, with the exception of one case with lower cross-helicity. However, the $f^{-1}$ regime is not apparent, as the correlation scale in this case is $\sim 5000$ s $(2\times10^{-4}$ Hz$)$.
On the other hand, the spectra of the SSW intervals exhibit a broad $f^{-5/3}$ regime extending to much lower frequencies with the $f^{-1}$ and $f^{-3/2}$ regimes being absent.

The variation of $K$ (defined in Section \ref{sec:methods}) as a function of $\tau$ is shown in Fig. \ref{fig8} for all the aforementioned intervals.
Similar to what has been observed during solar minima, $K$ is found to be scale dependent, decreasing with the time scale $\tau$ and approaching the Gaussian value $K\simeq1.67$ at $\tau>\sim10^4$ s (See Appendix \ref{AA}). 
This is again a clear indication of the non-universal nature of the distribution function of the magnetic field increments.
From Fig. \ref{fig8} (a), (b) \& (c), it is evident that a steeper power law followed by a shallower one is commonly observed for FSW and ASSW, while for SSW the shape varies quite a lot and it is hard to determine distinct regimes.
The two power law regimes existing for FSW and ASSW are fairly consistent with the existence of two distinct regimes in the spectra of these two types of wind (for more details see Section \ref{sec:solarminm}).
Moreover, for many intervals, the break in the kurtosis roughly corresponds to that in the power spectra between the $f^{-5/3}$ and $f^{-3/2}$ regimes in both FSW and ASSW, implying a close relation between them. 
For the SSW, where distinct regimes in kurtosis are not easily observed, the spectra is found to exhibit a broad $f^{-5/3}$ scaling.

The consolidated plot shown in Fig. \ref{fig8} (d) allows to perform a comparative study of intermittency among those three types of solar wind.
As evident from the plots, turbulence in ASSW is moderately intermittent, characterized by a value of $K$ which is intermediate between that of the SSW with the strongest intermittency and that of the FSW having the weakest intermittency.
Our observations are in agreement with the fact that, in the outer heliosphere, the SSW is in a state of more developed turbulence. 
This can also be inferred from the broader inertial range in the magnetic power spectra exhibited by SSW, extending to lower frequencies compared to FSW and ASSW.

We note that a recent study conducted by \cite{damicis2018} observes a $f^{-1}$ break in the spectra of ASSW at 1 au, clearly showing how the turbulence develops in ASSW by the broadening of the inertial range as it evolves with $R$.
Studies by \cite{damicis2015, damicis2018,damicis2021} explain the high  Alfv\'enicity of ASSW as due to its generation from coronal hole boundaries based on its composition and micro-physics. 

\section{SUMMARY AND CONCLUSION}
\label{sec:conclusion}

In this paper, we report the existence of two distinct sub-regimes for the inertial range in the magnetic power spectrum of solar wind turbulence within and beyond 1 au.
Although a single inertial range spectral power law has been traditionally observed \citep{BrunoCarbone2013}, a few studies have also identified variations in the spectral indices of the magnetic power spectrum \citep{Wicks2011,sioulas2023a} and in the scaling exponents of higher-order structure functions \citep{Wu2022,Wu2023,lsorriso2023}. 

In a series of fast, Alfvénic solar wind intervals measured during low solar activity, our findings show that a clear break in the kurtosis scaling closely coincides with the break observed in magnetic spectra separating the two sub-regimes characterized by $f^{-3/2}$ and $f^{-5/3}$ spectral power laws, both in the inner and outer heliosphere (see Fig. \ref{fig4}).
The appearance of a double power-law in both the magnetic power spectrum and kurtosis supports the robustness of this break, indicating the existence of a previously unidentified characteristic scale within the inertial range. 
Whereas the most probable explanation for the $f^{-5/3}$ regime can be obtained by the isotropic Kolmogorov phenomenology or anisotropic MHD turbulence with a weak ${\bf v}$-${\bf b}$ alignment in a non-Alfv\'{e}nic regime of solar wind fluctuations, the $f^{-3/2}$ regime could be reasonably associated with the anisotropic spectra along the strong ${\bf v}$-${\bf b}$ alignment \citep{Kolmogorov1941, goldreichsridhar1995, boldyrev2006}. 
Note that we consciously overlook the possibility of a $-3/2$ spectra by Iroshnikov-Kraichnan phenomenology \citep{kraichnan1965}, which is only valid for balanced MHD and cannot explain the emergence of $-3/2$ spectra when there is a strong ${\bf v}$-${\bf b}$ correlation.
A recent study by \cite{zhao2024} provided evidence of a transition from a weak to a strong turbulence regime as one moves from larger to smaller scales.
In our study, an inspection of the cross-helicity co-spectra revealed that the turbulence in FSW shifts from a highly imbalanced state ($|z^{+2}|\gg|z^{-2}|$, or vice-versa) at larger scales to a relatively balanced one ($|z^{+2}|\sim|z^{-2}|$) on moving towards the smaller scales (see Fig. \ref{fig6}).
These observations may explain the broken power-law behaviour of the spectrum and the kurtosis indicating a transition in the nature of turbulence as the cascade progresses towards the smaller scales.

We have further investigated the dependence of the sub-inertial regime break on the heliospheric distance, also in comparison with the ion and correlation scales. 
Our findings indicate a power-law behaviour for $l_K$ \citep[Taylor transformed $\tau_K$,][]{taylor1938} with $R$, which upon normalization with the typical ion scales (e.g. the ion-inertial scale $d_i$ and the ion gyro-radius $\rho_i$) and the correlation scale ($L_c$) practically disappears in the inner heliosphere (see Fig. \ref{fig7}). 
Therefore, both the correlation scale and the characteristic ion scale could be controlling the location of the break.
Interestingly, $l_K$ appears to approach the correlation scale shifting towards larger scales as $R$ increases, resulting in the absence of the $f^{-3/2}$ regime at $3.16$ au. 
This observation could explain the apparent transition of the inertial range magnetic spectral slope from $-3/2$ near the Sun to $-5/3$ farther away recently reported in the inner heliosphere \citep{chen2020, shi2021, Sioulas2023b}.
Note that a residual power-law radial dependence of the break scale still persists in the outer heliosphere, possibly due to variations in the latitude at which the FSW streams were sampled.
This residual behaviour of the normalized $l_K$ must be studied in depth in a future study as functions of the latitude and also the large-scale magnetic field angle, which may play a role in the degree of anisotropy in the measured turbulence.
Finally, we have performed a complementary preliminary analysis during high solar activity. 
Our results show similar but less robust features as described above for fast and slow solar wind, with single-scaling inertial ranges occasionally observed in FSW and ASSW streams. 
Due to the reduced sample used here, a possible role of latitudinal dependence cannot be ruled out.

The variability of wind types at solar maxima enables us to additionally characterize the state of turbulence in the Alfv\'enic slow solar wind, as compared to traditional fast and slow winds. 
ASSW, which is found in abundance near the ecliptic plane, is in an intermediate state of turbulence between typical fast and slow streams, with double scaling in the spectra but less regular scaling laws of the kurtosis. It is interesting that we do not observe the $f^{-1}$ break in the ASSW spectra \citep{mattheausgoldstein1986, chandran2018}, suggesting a possibility that this is located at a frequency lower than those accessible with our samples.
Interestingly, for the ecliptic solar wind at 1 au, the $f^{-1}$ break was found at the same frequency for both FSW and ASSW \cite{damicis2018}. In the current study, at distances greater than 1.5 au, similar $f^{-1}$ break in ASSW occurs at a much lower frequency compared to that for the FSW (see Fig. \ref{fig9}).

The observations described in this article clearly indicate the emergence of a characteristic scale within the inertial range of Alfvénic solar wind turbulence, likely separating two different dynamical regimes. 
The absence of double scaling in SSW at solar maximum seems to preliminarily suggest that Alfvénicity might play a crucial role in the occurrence of such break. 
However, a physical interpretation requires the disentanglement of several additional possible factors (e.g., geometry, expansion, latitude, turbulence amplitude), which is only possible through an extended statistical analysis, deferred to future works.
These findings could be of broad relevance for heliospheric studies, possibly informing solar wind modeling \citep[e.g.,][]{Cranmer2012,Usmanov2012,Zank2021,Chandran2021}, affecting energetic particle transport \citep{Pucci2016,Shukurov2017,Perri2021}, and more generally constraining the energetics of the solar wind \citep{Vasquez2007,rivera2024}.

\section*{Acknowledgments}

S.M. was supported by Students-Undergraduate Research Graduate Excellence (SURGE) summer internship program at Indian Institute of Technology Kanpur. 
S.B. acknowledges the financial support from the grant by Space Technology Cell-ISRO (STC/PHY/2023664O). 
L.S.-V. received support by the Swedish Research Council (VR) Research Grant N. 2022-03352 and by the International Space Science Institute (ISSI) in Bern, through ISSI International Team project \#23-591 (Evolution of Turbulence in the Expanding Solar Wind).

\section*{Data Availability}

For our study, we have used publicly available data from NASA CDAWeb (\href{https://cdaweb.gsfc.nasa.gov/}{https://cdaweb.gsfc.nasa.gov}) and AMDA science analysis system (\href{https://amda.irap.omp.eu/}{https://amda.irap.omp.eu}).


\appendix
\section{Estimation of kurtosis of magnetic field fluctuations following a Gaussian distribution}\label{AA}

Following the definition of the $n^{th}$ order structure function ($S_n$) given by eqn. (\ref{eqn1}), the expression of $S_4$ and $S_2$ takes the form:
\begin{equation}
    S_4 = ((\Delta B_r)^2 + (\Delta B_t)^2 + (\Delta B_n)^2)^2,
\end{equation}
and
\begin{equation}
    S_2 = ((\Delta B_r)^2 + (\Delta B_t)^2 + (\Delta B_n)^2),
\end{equation}
respectively. Now considering that the fluctuations follow a zero mean Gaussian distribution $f(\Delta B_i)$ having a standard deviation $\sigma$ such that
\begin{equation}
    f(\Delta B_i) = \frac{1}{\sqrt{2\pi\sigma^2}}exp\left[\frac{(\Delta B_i)^2}{2\sigma^2}\right],
\end{equation}
we have
\begin{equation}
    S_4 = \int\int\int ((\Delta B_r)^2 + (\Delta B_t)^2 + (\Delta B_n)^2)^2 f(\Delta B_r) f(\Delta B_t) f(\Delta B_n) \,d(\Delta B_r) \,d(\Delta B_t) \,d(\Delta B_n),
\end{equation}
and
\begin{equation}
    S_2 = \int\int\int ((\Delta B_r)^2 + (\Delta B_t)^2 + (\Delta B_n)^2) f(\Delta B_r) f(\Delta B_t) f(\Delta B_n) \,d(\Delta B_r) \,d(\Delta B_t) \,d(\Delta B_n),
\end{equation}
which seem a bit rigorous but can be solved easily to obtain $S_4 = 15\sigma^4$ and $S_2=3\sigma^2$. Thus, the kurtosis defined by eqn. (\ref{eqn2}) takes the value $K=5/3 \simeq 1.67$.

\section{Component-wise kurtosis of the magnetic field fluctuations in FSW intervals during solar minimum}\label{AB}

In this appendix we show the kurtosis $K$ for each individual RTN magnetic field component, using four example intervals from both Helios and Ulysses database, at eight different distances from the Sun. Different colours refer to different intervals. 
Whenever present, a power law is shown as coloured dashed line, and the corresponding scaling exponents are indicated in each panel. 
Two power laws can be identified in all of the Helios and most of the Ulysses intervals, with the exception of the radial component at 2.75 au and of all components at 3.16 au. 
The timescale $\tau_K$ of the break between the two power laws is indicated by a solid vertical grey line, while the dashed grey vertical lines indicate the location of the spectral break, $1/f_b$.

\begin{figure}[h]
    \centering
    \includegraphics[width=1\linewidth]{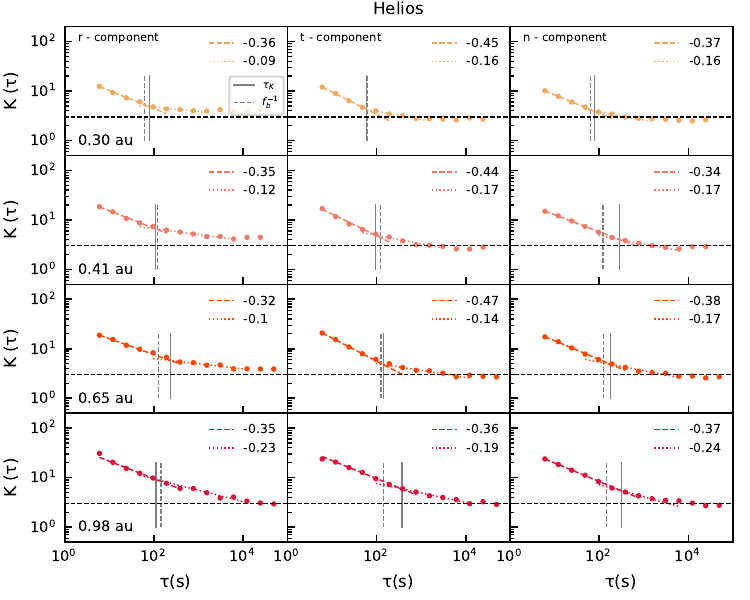}
    \caption{Component-wise kurtosis $K(\tau)$ of magnetic field fluctuations for several intervals of FSW during periods of solar minima for Helios data (year 1976) in the inner heliosphere from a sustained coronal hole. The three columns represent the r, t and n components of $K$. Power-law fits and the corresponding scaling exponents are indicated. Vertical lines indicate the observed break, $\tau_K$ (solid), and the timescale corresponding to the spectral break, $1/f_b$ (dashed).}
    \label{fig:app1}
\end{figure}
\begin{figure}[h]
    \centering
    \includegraphics[width=1\linewidth]{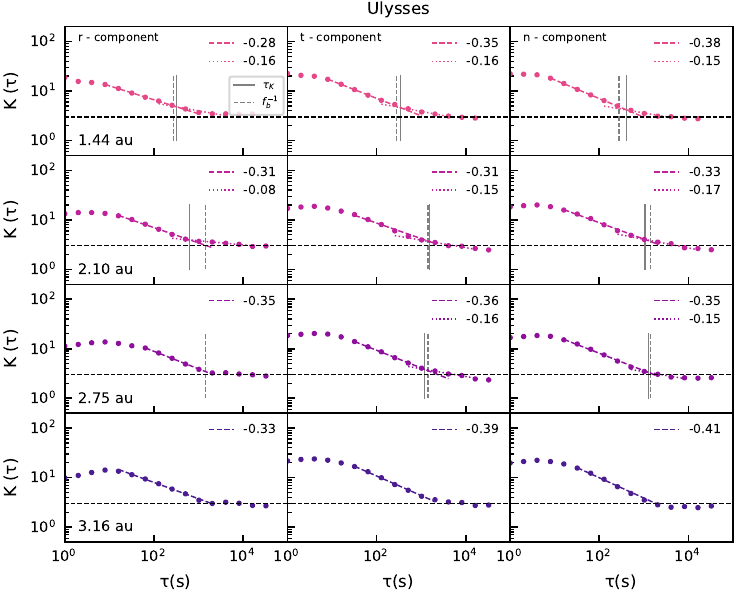}
    \caption{Component-wise kurtosis $K(\tau)$ of magnetic field fluctuations for several intervals of FSW during periods of solar minima for Ulysses data (years 1995-1996) in the outer heliosphere at varying distances and latitudes. The three columns represent the r, t and n components of $K$. Power-law fits and the corresponding scaling exponents are indicated. Vertical lines indicate the observed break, $\tau_K$ (solid), and the timescale corresponding to the spectral break, $1/f_b$ (dashed).}
    \label{fig:app2}
\end{figure}
\newpage

\nocite*
\bibliography{draft}{}
\bibliographystyle{aasjournal}



\end{document}